%% file: main.tex
\documentclass[preprint,copyright,creativecommons]{eptcs}
\providecommand{\event}{PLACES 2016} 
\usepackage{hyperref}
\usepackage{amsmath}
\usepackage{listings}
\usepackage{latexsym}
\usepackage{amsthm}
\usepackage{xspace}
\usepackage{soul} 
\usepackage[capitalise,noabbrev]{cleveref}
\usepackage{tikz}
\usetikzlibrary{decorations.pathreplacing}
\usetikzlibrary{matrix}
\usetikzlibrary{calc}
\usepackage{subcaption}
\usepackage{xcolor}

\definecolor{keywords}{rgb}{0.13,0.13,1}
\definecolor{comments}{rgb}{0,0.5,0}
\definecolor{strings}{rgb}{0.9,0,0}
\definecolor{dkgreen}{rgb}{0,0.5,0}
\definecolor{dkviolet}{rgb}{0.58, 0.0, 0.83}
\definecolor{ltblue}{rgb}{0.36, 0.54, 0.66}

\usepackage{lstcoq}
\input{macros}

\input{macros-lang}

\input{macros-codestyle}

\author{
Tiago Cogumbreiro
\institute{Rice University}
\and
Jun Shirako
\institute{Rice University}
\and
Vivek Sarkar
\institute{Rice University}
}
\def\titlerunning{Formalization of Phase Ordering}
\title{\titlerunning}
\def\authorrunning{T. Cogumbreiro, J. Shirako, V. Sarkar}

\begin{document}
\maketitle
\input{abstract}
\input{introduction}
\input{semantics}
\input{phase-ordering}
\input{conclusion}

\bibliographystyle{eptcs}
\bibliography{refs}
\newpage
\appendix
\input{coq-results}
\input{proofs}

\end{document}

%% file: macros.tex
\newtheorem{definition}{Definition}
\newtheorem{theorem}{Theorem}
\newtheorem{lemma}{Lemma}
\newtheorem{remark}{Remark}
\newtheorem{example}{Example}

\newcommand{\ie}{\textit{i.e.,}\@\xspace}
\newcommand{\eg}{\textit{e.g.,}\@\xspace}
\newcommand{\etal}{\textit{et al.}\@\xspace}

%% file: macros-lang.tex
\newcommand{\Osignal}{\mathtt{signal}}
\newcommand{\Owait}{\mathtt{wait}}
\newcommand{\Fregister}{\mathtt{reg}}
\newcommand{\Oregister}[2][\Tid]{\Fregister({#1},{#2})}
\newcommand{\Odrop}{\mathtt{drop}}

\newcommand{\reduces}[1][]{\xrightarrow{{#1}}}

\newcommand{\Reduces}[2][\Tid]{\reduces[{#1}:{#2}]}
\newcommand{\Rsignal}[1][\Tid]{\Reduces[{#1}]\Osignal}
\newcommand{\Rwait}[1][\Tid]{\Reduces[{#1}]\Owait}
\newcommand{\Rregister}[3][\Tid]{\Reduces[{#1}]{\Oregister[{#2}]{#3}}}
\newcommand{\Rdrop}[1][\Tid]{\Reduces[{#1}]\Odrop}

\newcommand{\Tid}{t}
\newcommand{\Tidy}{t'}
\newcommand{\TidDom}{\mathcal T}
\newcommand{\ViewDom}{\mathcal V}
\newcommand{\ViewWF}{{\mathcal V}^{WF}}
\newcommand{\PhDom}{\mathcal P}
\newcommand{\PhWO}{{\mathcal P}^{WO}}
\newcommand{\PhWF}{{\mathcal P}^{WF}}
\newcommand{\View}{v}
\newcommand{\Viewy}{w}
\newcommand{\Viewz}{u}
\newcommand{\Ph}{P}
\newcommand{\Phy}{Q}
\newcommand{\Phz}{R}
\newcommand{\Op}{o}

\newcommand{\SO}{\mathtt{SO}}
\newcommand{\WO}{\mathtt{WO}}
\newcommand{\SW}{\mathtt{SW}}

\newcommand{\tv}[3][\Nat]{\{\FsigPhase\assign{#1},\FwaitPhase\assign{#2},\Fmode\assign{#3}\}}

\newcommand{\update}[2]{[{#1}\mapsto{#2}]}
\newcommand{\Fmode}{\texttt{mode}}
\newcommand{\FwaitPhase}{\texttt{wp}}
\newcommand{\FsigPhase}{\texttt{sp}}
\newcommand{\WaitPhase}[1][\View]{{#1}.\FwaitPhase}
\newcommand{\SigPhase}[1][\View]{{#1}.\FsigPhase}
\newcommand{\Mode}[1][\View]{{#1}.\Fmode}

\newcommand{\CanWait}[1][\View]{\mathrm{Waiter}\ {#1}}
\newcommand{\CanSignal}[1][\View]{\mathrm{Signaler}\ {#1}}

\newcommand{\Nat}{n}
\newcommand{\Naty}{m}

\newcommand{\Sync}[2][\Tid]{{#1}\ \mathrm{sync}\ {#2}}
\newcommand{\Await}[2][\Ph]{\mathrm{Await}({#1},{#2})}
\newcommand{\assign}{\mathbin{\texttt{:=}}}

\newcommand{\Regmode}{r}

\newcommand{\HB}{\prec}
\newcommand{\CHB}{\unrhd}
\newcommand{\PAR}{\mathbin{||}}

\newcommand{\pad}{\;\;}
\newcommand{\Space}[1]{\pad{#1}\pad}
\newcommand{\Grmeq}{{::=}}
\newcommand{\grmeq}{\Space\Grmeq}

\newcommand{\Grmor}{\mid}
\newcommand{\grmor}{\;\Grmor\;}

\newcommand{\eqdef}{\overset{\mathrm{def}}{=}}
\newcommand{\dom}{\mathrm{dom}}

%% file: macros-codestyle.tex
\usepackage{color}
\usepackage{listings}

\lstdefinelanguage{hj}{
  language     = Java,
  morekeywords = {signal,doWait,newPhaser,asyncPhased,drop},
}

\lstnewenvironment{code}[1][]
  {%
  \lstset{language=hj,#1}%
}
{}

%% file: abstract.tex
\begin{abstract}
  Phasers pose an interesting synchronization mechanism that
  generalizes many collective synchronization patterns seen in
  parallel programming languages, including barriers, clocks, and
  point-to-point synchronization using latches or semaphores.
  This work characterizes scheduling constraints on phaser operations,
  by relating the execution state of two tasks that operate on the
  same phaser.
  We propose a formalization of Habanero phasers,
  May-Happen-In-Parallel, and Happens-Before relations for phaser
  operations, and show that these relations conform with the semantics.
  Our formalization and proofs are fully mechanized using the Coq
  proof assistant, and are available online.
\end{abstract}

%% file: introduction.tex
\section{Introduction}

Phasers are an interesting synchronization mechanism that generalizes
barriers with collective producer-consumer synchronization.
A phaser can encode the synchronization mechanism of latches, futures,
join barriers, cyclic barriers, as well as any \emph{collective
  synchronization pattern} provided by CUDA, C$\sharp$, Java, MPI, and
X10.
Phasers~\cite{shirako.peixotto.etal:phasers} were first introduced in
the Habanero Extreme Scale research project at Rice University, as an
extension to X10 clocks~\cite{charles.etal:x10}, and implemented in
Habanero-Java and Habanero-C.
A restricted form of phasers was also introduced in the standard
\texttt{java.util.concurrent.Phaser} library starting with Java~7.
The phaser synchronization mechanism is relevant at the
theoretical level because of its generality.
Theoretical results that target phasers can easily translate across
different languages and parallel runtimes~\cite{cogumbreiro:armus}.

The phaser synchronization mechanism lets tasks observe a collective
event, called \emph{phase}, which is visible once every member of a
group of tasks \emph{signals} the phaser exactly once.
We define \emph{signalers} of the phaser as the group of tasks able to
signal a phaser.
The same phaser can be used to observe multiple phases, which
are distinguishable by a natural number.
A task can observe phase~$n$ once each signaler issues at least~$n$
signals.
Phaser synchronization also features \emph{dynamic membership}, that
is, the group of signalers can grow and shrink dynamically: a signaler
can add a member, which in turn inherits the signal count of the task
adding it;
a signaler can also revoke its membership at any time.

As an example of phaser synchronization, let us consider a group of
three tasks, uniquely identified by~$\Tid_1$, $\Tid_2$, and $\Tid_3$,
and let this group of tasks be the signalers of phaser~$\Ph$.
Also, lets examine a point in time, with respect to phaser~$\Ph$,
where task~$\Tid_1$ signaled 3 times, task~$\Tid_2$ signaled 4 times,
and task~$\Tid_3$ signaled 10 times.
Tasks can use~$\Ph$ to observe any phase below or equal to phase~$3$,
since the signalers collectively issued at least 3 signals.
Conversely, at this point in time, any phase above~$3$ is \emph{not}
observable, \eg for phase~4 to be observed we are missing a signal
from task~$\Tid_1$.
Dynamic membership affects synchronization:
if task~$\Tid_1$ adds a task~$\Tid_4$ as a signaler of phaser~$\Ph$,
then for phase~4 to be observable we are missing a signal from
task~$\Tid_1$ and a signal from task~$\Tid_4$;
and, if, subsequently, tasks~$\Tid_1$ and~$\Tid_4$ revoke their
membership, then phase~$4$ is observable.

This paper introduces the first formalization of Habanero phasers and
also presents an Happens-Before (HB)~\cite{lamport:time-clocks}
relation and a May-Happen-In-Parallel~\cite{duesterwald:conc-analysis}
(MHP) relation for phaser operations, both of which are fundamental
problems for concurrency analysis.
MHP and HB characterize scheduling restrictions between two
instruction instances.
An example of the HB relation is ordering any instruction that
happened before spawning a task and any instruction in the task body
being spawned; MHP can be defined using the HB relation.
MHP and HB analysis are fundamental in the verification of barrier
synchronization errors~\cite{sharma:prod-cons-gpu}, lock-based
deadlock prediction~\cite{cai:conlock,elmas:goldilocks}, and
race-detection~\cite{maiya:android-race,
  smaragdakis:race-detection-poly}.

\input{fig-prog1}

The HB relation we introduce comes from the Phase Ordering
definition~\cite{shirako.peixotto.etal:phasers} that relates the
execution state of two tasks manipulating the same phaser: if the
number of signals issued by a task (property~\texttt{sp}) is smaller
than the last phase observed by some other task (property~\texttt{wp})
then the former task happened before the latter task.
The execution of a program that uses Habanero phasers must respect the
scheduling restriction imposed by Phase Ordering, but how can we be
sure that this property holds?
The example in \cref{fig:prog1} lists the execution trace of two
tasks and also includes the~\texttt{sp} and~\texttt{wp} at each step,
for both tasks, and w.r.t.\@ the same phaser.
Tasks increment their~\texttt{sp} after signaling, and their~\texttt{wp}
after waiting.
Note how HB orders instructions at different points in time:
the assignment \lstinline|x:=1| by~$\Tid_1$ happens before
\lstinline|println(x)| by~$\Tid_2$, so we can conclude that $\Tid_2$
reads the value written by task~$\Tid_1$.
Conversely, HB does not order the write \lstinline|y:=2| by $\Tid_2$
and the read~\lstinline|println(y)| by~$\Tid_1$, so, as read and write
are unsynchronized, there is a data race.
The goal of this work is twofold: 1) prove that to schedule
phaser-operation across tasks it is sufficient to compare a task-local
property~\texttt{sp} of one task with the last global
observation~\texttt{wp} of another task; and 2) and prove that the
semantics of Habanero phasers respects the Phase Ordering property.

Crafa \etal propose a Coq formalization of a subset of the X10 and
define a HB relation in~\cite{crafa:semantics-res-x10}, but only
consider fork-join synchronization, and omit dynamic barrier
synchronization that we formalize.
%
%
Tomofumi \etal use HB and MHP to check for data races in the
polyhedral subset of clocked X10
programs~\cite{tomofumi:clock-race-freedom}.
Joshi \etal propose an informal MHP relation for X10
clocks~\cite{joshi:mhp-dyn-barriers}.

This paper establishes two main properties with respect to the phasers
semantics we introduce.
First, as required by HB and MHP analysis, we show that the HB
relation we define is a causality relation~\cite{lamport:time-clocks}.
Second, since the HB relation is defined on the state of a
phaser~$\Ph$, we show that HB \emph{conforms} with the execution
semantics of phasers; that is, if a state~$\Ph$ reduces to~$\Phy$
after zero or more steps, then $\Phy$ cannot happen before~$\Ph$.
By targeting phasers, our formalization unifies collective
producer-consumer synchronization~\cite{sharma:prod-cons-gpu} and
barriers with dynamic membership~\cite{joshi:mhp-dyn-barriers} in a
single theoretical framework.
Additionally, we formalize and establish the correctness of our
definitions with proofs verified by the Coq proof assistant, available
online, as part of our \textbf{HJ-Coq formalization project~\cite{hj-coq}}.

The main contributions of this paper are:
\begin{enumerate}
\item introduces the first formalization semantics of Habanero phasers;
\item defines an HB relation and an MHP relation for phaser operations;
\item shows that HB is a causality relation, given by \cref{lem:wo-pres-red};
\item shows that HB conforms with the reduction relation, given by \cref{thm:hb-sred}; 
\item presents the full Coq mechanization of the theory, along with examples.
\end{enumerate}

In the next section, we describe the phaser operations and its  semantics.
In \cref{sec:phase-ordering}, we introduce Phase Ordering, the MHP
relation, and the HB relation.
Next, in \cref{sec:results}, we establish the main results with
regards to the semantics of phasers.
We conclude in \cref{sec:conclusion} and discuss future directions.

%% file: fig-prog1.tex
\begin{figure}
  \centering
\scalebox{0.76}{
\begin{tikzpicture}[ampersand replacement=\&,
nodes={minimum width=6em,minimum height=1.6em}]
\matrix (m) [matrix of nodes]
{
{}  \&          $\Tid_1$ \&  $\Tid_2$ \& {} \\
\texttt{sp:0,wp:0}  \& |[fill=green!20,rounded corners]|  x:=1     \&  $\Osignal$ \& \texttt{sp:0,wp:0} \\
\texttt{sp:0,wp:0}  \&        $\Osignal$ \& |[fill=red!20,rounded corners]| y:=2 \& \texttt{sp:1,wp:0}\\
\texttt{sp:1,wp:0}  \&        $\Owait$   \& $\Owait$ \& \texttt{sp:1,wp:0} \\
\texttt{sp:1,wp:1}         \& |[fill=red!20,rounded corners]|  println(y)     \& |[fill=green!20,rounded corners]| println(x) \& \texttt{sp:1,wp:1}\\
};
\draw (m-1-1.south west) -- (m-1-4.south east);
\draw (m-1-3.north west) -- (m-5-3.south west);

    \path[thick,->] ($ (m-2-2.center) + (18pt,0)$) edge [bend left] ($ (m-5-3.center) - (12pt,0)$);
    \path[thick,->] ($ (m-2-3.center) - (20pt,0)$) edge [bend right] ($ (m-5-2.center) + (16pt,2pt)$);
  \end{tikzpicture}
}
  \caption{Phase-ordering between two task traces in a program with a race error.}
  \label{fig:prog1}
\end{figure}
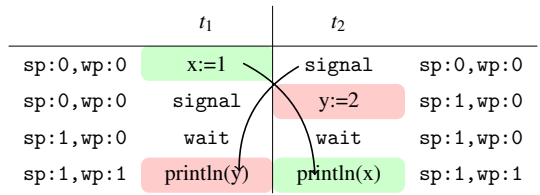

%% file: semantics.tex
\section{Phaser semantics}
\label{sec:semantics}

Let us discuss informally the semantics of Habanero phasers by
revisiting the example in \cref{fig:prog1}.
In the following Java code listing, tasks~$\Tid_1$ and~$\Tid_2$
synchronize each of their access to two different shared
variables~\lstinline{x} and~\lstinline{y} by means of a
phaser~\lstinline{ph}.

\begin{code}[numbers=left,xleftmargin=\parindent]
ph = newPhaser(SIG_WAIT);                 (*@\label{l:t1-new-ph}@*)
asyncPhased(ph.inMode(SIG_WAIT), () -> {  (*@\label{l:t1-async}@*)
  ph.signal();                            (*@\label{l:t2-signal}@*)
  y = 2;                                               (*@\label{l:t2-write}@*)
  ph.doWait();                            (*@\label{l:t2-wait}@*)
  println(x); (*@\label{l:t2-read}@*)
  ph.drop(); (*@\label{l:t2-drop}@*)
});
x = 1; (*@\label{l:t1-write}@*)
ph.signal();                              (*@\label{l:t1-signal}@*)
ph.doWait();                              (*@\label{l:t1-wait}@*)
println(y); (*@\label{l:t1-read}@*)
\end{code}

Task~$\Tid_1$ creates phaser~\lstinline{ph} in \cref{l:t1-new-ph}
and then spawns a task~$\Tid_2$ in \cref{l:t1-async}.
These two tasks are the signalers of~\lstinline|ph|.
The creator of a task is a signaler of that phaser.
Signalers, and only signalers, can register other members by spawning them
with~\lstinline|asyncPhased| and passing the target phaser.
Here, task~$\Tid_1$ registers task~$\Tid_2$ with phaser~\lstinline|ph|
--- we postpone discussing the meaning of
expression~\lstinline|ph.inMode(SIG_WAIT)| in \cref{l:t1-async};
for now it is enough to interpret the expression as~\lstinline|ph|.
Task~$\Tid_1$ then writes to variable~\lstinline{x} in
\cref{l:t1-write} and reads from variable~\lstinline{y} in
\cref{l:t1-read}, while, concurrently, task~$\Tid_2$ writes to
variable~\lstinline{y} in \cref{l:t2-write} and reads from
variable~\lstinline{x} in \cref{l:t2-read}.
Before terminating, task~$\Tid_2$ revokes its membership on
phaser~\lstinline|ph| by invoking~\lstinline|ph.drop()| in
\cref{l:t2-drop}.

Tasks~$\Tid_1$ and~$\Tid_2$ synchronize in the example by
executing~\lstinline|ph.doWait()| in \cref{l:t2-wait,l:t1-wait}: each
task waits for phase~1 to be observed, which can only happen once both
task execute \lstinline|ph.signal()| in \cref{l:t2-signal,l:t1-signal}.
The reason there is a data race in \cref{l:t2-write,l:t1-read} is
because task~$\Tid_1$, that blocks with~\lstinline|ph.doWait()| before
reading~\lstinline|y|, can unblock and read the variable when
task~$\Tid_2$ signals in \cref{l:t2-signal}.
But since task~$\Tid_2$ writes to~\lstinline|y| \emph{after}
signaling, then the read from task~$\Tid_1$ in \cref{l:t1-read} runs
concurrently with the write of task~$\Tid_2$ in \cref{l:t2-write}.

A feature that distinguishes Habanero phasers from other barrier-like
synchronization mechanisms is that waiting for signalers is
\emph{optional}.
A task can choose to manipulate a phaser according to two abilities:
(i) the ability to observe phaser synchronization, \ie waiter, and (ii)
the ability to influence synchronization, \ie signaler.
Each member is registered according to a mode~$\Regmode$ among: $\SW$
for tasks that must signal and wait, $\WO$ for tasks that wait
but do not signal, and $\SO$ for tasks that signal but do not wait.
In the example, task~$\Tid_1$ registers task~$\Tid_2$ in
phaser~\lstinline|ph| using mode $\SW$, which is short
for~\lstinline|SIG_WAIT|, given by expression
\lstinline|ph.inMode(SIG_WAIT)| in \cref{l:t1-async}.

%
Waiting observes the signals from \emph{every} signaler.
Thus, tasks that wait and signal, mode~$\SW$, as in the example, must
signal before waiting at every phase to prevent waiting for a signal
the task did not produce.
This synchronization pattern is known as a \emph{barrier}.
Members disregard wait-only ($\WO$) tasks upon waiting; this subset of
tasks cannot influence synchronization, only observe it.
Phasers can encode latches, future-promises, and fork-join synchronization
patterns using wait-only tasks.
Finally, tasks that only signal, do not wait for others; this lets
phasers encode producer-consumer synchronization.
%
%

\paragraph{HJ Phaser formalization.}
We define the state of a phaser~$\Ph$ to be a map from
members~$\TidDom$ into \emph{views}~$\ViewDom$, which holds the signal
count, the wait count, and registration mode of a member.
Consider the usual operations on finite maps (which we use to encode
phasers) with the given notation: predicate $\Ph(\Tid) = \View$
ensures that the pair of key~$\Tid$ and value~$\View$ is a member of
map~$\Ph$, predicate~$\Tid \in \Ph$ is short-hand
for~$\exists\View\colon\Ph(\Tid) = \View$, map~$\Ph\update \Tid \View$
adds the pair $\Tid$ and~$\View$ to map~$\Ph$ (replacing the assigned
view if~$\Tid \in \Ph$), and map~$\Ph - \Tid$ results from removing
the pair associated with key~$\Tid$ from map $\Ph$.
In the mechanization, we use Coq's standard library of finite maps
\href{https://coq.inria.fr/library/Coq.FSets.FMaps.html}{Coq.FSets.FMaps}.

A view~$\View$ represents the task-local information that each
member has over the phaser.
The view consists of a triple: the first value~$\Nat$ is a natural
number that counts the number of times the given task issued a signal
on the target phaser and can be accessed by~$\SigPhase$; the second
value~$\Naty$ counts the number of waits and can be accessed
by~$\WaitPhase$; the third value~$\Regmode$ is the registration mode
of the given task and is accessed by~$\Mode$.
$$
\View \grmeq \tv \Naty \Regmode
$$
The field update operation~$\View.f \assign e$ yields a view that is
the same as~$\View$ except for field~$f$ that becomes~$e$.
For instance,~$\SigPhase \assign 3$ yields a view, say~$\Viewy$,
where~$\WaitPhase[\Viewy] = \WaitPhase$, $\SigPhase[\Viewy]=3$, and
$\Mode[\Viewy] = \Mode$.
To inquire the signaling and waiting abilities of a view we have the
following predicates:
$\CanWait[\Regmode] \eqdef \Regmode \in \{\WO, \SW\}$,
$\CanSignal[\Regmode] \eqdef \Regmode \in \{\SO, \SW\}$.
And let the short-hand notation $\CanSignal \eqdef \CanSignal[\Mode]$
and $\CanWait \eqdef \CanWait[\Mode]$.

\input{fig-semantics}

We define a small-step operational semantics for phaser operations in
\cref{fig:semantics}.
The reduction~$\Reduces \Op$ is labeled by the member~$\Tid$ issuing
the operation, and by an operation~$\Op$ defined below.
$$
\Op \grmeq \Osignal \grmor \Owait \grmor \Oregister \Regmode\grmor \Odrop
$$

%
\begin{remark}
  To model Java phasers and X10 clocks semantics refer to
  \cref{fig:semantics} but limit the registration mode to
  signal-wait mode, that is $\Regmode \grmeq \SW$.
\end{remark}

Operation~$\Osignal$ increments the signal phase.
Only tasks registered as signalers can issue this operation. 
The pre-conditions in signal, $\WaitPhase = \SigPhase$, and in wait,
$\WaitPhase + 1 = \SigPhase$, enforce tasks registered in signal-wait
mode to interleave each signal with a wait.
%

Waiting is the crux of synchronization; this is captured
by~$\Sync \Ph$, defined next.

$$
\frac{
  \Mode[\Ph(\Tid)] = \SO
}{
  \Sync \Ph
}
\qquad
\frac{
  \CanWait[\Ph(\Tid)] 
  \quad
  \Await {\WaitPhase[{\Ph(\Tid)}] + 1}
}{
  \Sync \Ph
}
$$

Signal-only tasks do not wait for others, so $\Sync \Ph$ holds
in this case.
Waiter task~$\Tid$ must await the subsequent
wait-phase~$\WaitPhase[\Ph(\Tid)] + 1$.
Proposition~$\Await \Nat$ holds once phase~$\Nat$ can be observed; the
definition ensures that all tasks that can signal have issued at
least~$\Nat$ signals.
$$
\Await \Nat \eqdef \forall \Tid \colon\
  \CanSignal[\Ph(\Tid)]
\implies
  \SigPhase[ \Ph(\Tid) ] \ge \Nat
$$

Tasks register other tasks with with \lstinline|asyncPhased|, which is
captured by~$\Oregister \Regmode$.
For instance, instruction
\lstinline|asyncPhased(ph.inMode(SIG_WAIT),...)| in
\cref{l:t1-new-ph} becomes~$\Oregister[\Tid_2] \SW$ in this semantics
if we are spawning task~$\Tid_2$.
Habanero phasers limit task registration: only unregistered tasks can
be added, thus $\Tidy \in \Ph$, only registered tasks~$\Tid$ can add
new members, $\Ph(\Tid) = \View$, only waiters can register other
waiters, hence $\CanWait [\Regmode] \implies \CanWait$, and only
signalers can register other signalers, so
$\CanSignal [\Regmode] \implies \CanSignal$.

To establish the results in the next section, let us establish the
invariant of well-formedness.
Additionally, let~$\Ph \reduces\Phy$ be defined as there exist $\Tid$
and $\Op$ such that~$\Ph \Reduces \Op \Phy$.

\begin{definition}[Well-formed view]
  \label{def:wf-views}
  Let a well-formed view be such that~$\WaitPhase \le \SigPhase$ and if
  $\CanWait$ then $\SigPhase - \WaitPhase \le 1$.
  Let~$\ViewWF$ be the set of all well-formed views.
\end{definition}

\begin{lemma}[Reduction preserves well-formedness of views]
  \label{lem:red-pres-wf}
  Let~$\Ph$ be such that if $\Ph(\Tid) = \View$, then
  $\View \in \ViewWF$.
  If $\Ph \reduces \Phy$, then $\Phy$ is such that if
  $\Phy(\Tid) = \View$, then $\View \in \ViewWF$.
\end{lemma}

Henceforth, we only consider views that are in~$\ViewWF$.

%% file: fig-semantics.tex
\begin{figure}
  \centering
  \begin{gather*}
    \frac{
      \Ph(\Tid) = \View
      \qquad
      \CanSignal
      \qquad
      \Mode = \SW \implies \WaitPhase = \SigPhase
    }{
      \Ph \Rsignal \Ph \update \Tid {\SigPhase \assign \SigPhase + 1}
    }
    \\
    \frac{
      \Sync \Ph
      \qquad
      \Ph(\Tid) = \View
      \qquad
      \CanWait
      \qquad
      \Mode = \SW \implies \WaitPhase + 1 = \SigPhase
    }{
      \Ph
      \Rwait
      \Ph \update \Tid {\WaitPhase \assign \WaitPhase + 1}
    }
    \\
    \frac{
       \Tidy \notin \Ph
       \qquad
       \Ph(\Tid) = \View
       \qquad
       \CanWait [\Regmode] \implies \CanWait
       \qquad
       \CanSignal [\Regmode] \implies \CanSignal
    }{
      \Ph {\Rregister \Tidy \Regmode} \Ph \update {\Tidy} {\Mode \assign \Regmode}
    }
    \\
    \frac{
      \Tid \in \Ph
    }{
      \Ph
      \Rdrop
      \Ph - \Tid
    }
  \end{gather*}
  \caption{Operational semantics of phaser operations}
  \label{fig:semantics}
\end{figure}

%% file: phase-ordering.tex
\section{Phase Ordering}
\label{sec:phase-ordering}

This section formalizes Phase Ordering, originally introduced
in~\cite{shirako.peixotto.etal:phasers}, to reason about whether two
tasks should execute concurrently in terms of views~$\ViewDom$ and
states~$\PhDom$.
Specifically, Phaser Ordering is a Happens-Before relation: if the
number of signals issued by a task is smaller than the last phase
observed by some other task, then the former Happened Before the
latter.
%
%
For instance, let~$\View_1$ be a view that task~$\Tid_1$ has over the
phaser in \cref{fig:prog1} and $\View_2$ be a view that task~$\Tid_2$
has over the phaser in \cref{fig:prog1}, each from a distinct state of
the same phaser~\lstinline|ph|, \ie there exists two states~$\Ph$
and~$\Phy$ such that $\Ph(\Tid_1) = \View_1$ and
$\Phy(\Tid_2) = \View_2$.
Now,
let~$\View_1 \eqdef \{\FsigPhase\assign 0, \FwaitPhase \assign 0,
\Fmode \assign \SW \}$
be the view of~$\Tid_1$ when executing \lstinline|x:=1| and
$\View_2 \eqdef \{\FsigPhase \assign 1, \FwaitPhase \assign 1, \Fmode
\assign \SW \} $
be the view~$\Tid_2$ when executing \lstinline|println(y)|.
The registration mode tells us that view~$\View_2$ must observe and
wait for the signals of~$\View_1$.
View~$\View_1$ tells us that $\Tid_1$ did not produce any signal and
$\View_2$ tells us that~$\Tid_2$ observed phase~1 (a collective
signal).
Thus, since~$\Tid_1$ signaled fewer times than the phase observed
by~$\Tid_2$, we can infer that~$\View_1$ must have happened
before~$\View_2$.
In order for view~$\View_2$ to observe a signal, the task controlling
view~$\View_1$ must eventually signal to
become~$\SigPhase[\View_1]=1$.
%

\begin{definition}[Happens-before (HB) relation]
  \label{def:hb}
  Let $\View_1 \HB \View_2 $ read as $\View_1$ must have happened before
  $\View_2$, defined as the conjunction of:
$$
\CanSignal[\View_1]
\qquad
\SigPhase[\View_1] < \WaitPhase[\View_2]
\qquad
\CanWait[\View_2]
$$
We say that $\Ph \HB \Phy$ if there exist two tasks $\Tid, \Tidy$ such that
$\Ph(\Tid) \HB \Phy(\Tidy)$.
\end{definition}

\begin{example}
  \label{ex:hb}
  Suppose~$\Ph \Rsignal \Phy \Rwait \Phz$ and that $\Mode[\Ph(\Tid)] = \SW$.
  We have that~$\Ph \HB \Phz$.
\end{example}
\begin{proof}
  Let~$\Ph(\Tid) = \View$, $\Phy(\Tid) = \Viewy$, and $\Phz(\Tid) = \Viewz$.
  First, we simplify our goal, since we know that from $\Owait$
  $\WaitPhase[\Viewz] = \WaitPhase[\Viewy] + 1$ and because~$\Osignal$
  does not alter the wait phase we have that
  $\WaitPhase[\Viewy] = \WaitPhase$.
  Hence,
  $ \SigPhase < \WaitPhase[\Viewz] \equiv \SigPhase <
  \WaitPhase[\Viewy] + 1 \equiv \SigPhase < \WaitPhase + 1 $.
  Now, by inverting reduction~$\Rsignal$, we get two cases: either
  $\WaitPhase = \SigPhase$ and it trivially holds, otherwise we get a
  contradiction.
\end{proof}

Let us show that $\HB$ is a causality relation over views,
a fundamental notion for many problems occurring in distributed
computing~\cite{schwarz:detecting-causal-rel}.
By causality relation we mean a strict partial order: (i)
\emph{transitive:} if $\View_1 \HB \View_2$ and $\View_2 \HB \View_3$,
then $\View_1 \HB \View_3$; (ii) \emph{irreflexive:} for all $\View$,
we have that~$\neg (\View \HB \View)$; (iii)~\emph{asymmetric:} if
$\View_1 \HB \View_2$, then $\neg (\View_2 \HB \View_1)$.

\begin{lemma}
  \label{lem:hb-caus}
  $(\HB,\ViewDom)$ is a causality relation.
\end{lemma}

To show that $\HB$ is a causality relation over phasers, we need to
establish some auxiliary results that reason about states that cannot
happen before others, $\neg (\Ph \HB \Phy)$.
Since Coq uses a constructive logic, it is easier avoid the use of
false in our premises.
Let the negation of Happens-Before be defined as Cannot-Happen-Before.

\begin{definition}[Cannot-Happen-Before relation]
  \label{def:chb}
  Let $\View_1 \CHB \View_2$ read as $\View_1$ cannot happen before~$\View_2$:
$$
\Mode[\View_1] = \WO
\quad
\vee
\quad
\SigPhase[\View_1] \ge \WaitPhase[\View_2]
\quad
\vee
\quad
\Mode[\View_2] = \SO
$$
We say that $\Ph \CHB \Phy$ if for any tasks $\Tid, \Tidy$ we have that
$\Ph(\Tid) \CHB \Phy(\Tidy)$.
\end{definition}

Although our proofs use $\CHB$, the following remark allow us to
present our lemmas with the more familiar HB relation,
$\neg (\Ph \HB \Phy)$.
\begin{remark}
  \label{rem:conv}
  We have  that
  $\View_1 \HB \View_2 \iff \neg (\View_1 \CHB \View_2)$,
  $\View_1 \CHB \View_2 \iff \neg (\View_1 \HB \View_2)$,
  $\Ph_1 \HB \Ph_2 \implies \neg (\Ph_1 \CHB \Ph_2)$, and
  $\Ph_1 \CHB \Ph_2 \implies \neg (\Ph_1 \HB \Ph_2)$
\end{remark}

Finally, we define the usual notion of May-Happen-in-Parallel (or
concurrency relation) for views and for phasers.
\begin{definition}[May-Happen-in-Parallel relation]
  Let $\View_1 \PAR \View_2 $ read as $\View_1$ happens in parallel
  with $\View_2$ and be defined as $\View_1 \CHB \View_2$ and
  $\View_2 \CHB \View_1$.
  Let $\Ph_1 \PAR \Ph_2 $ be defined as $\Ph_1 \CHB \Ph_2 $ and $\Ph_2 \CHB \Ph_1$.
\end{definition}

\section{Results}
\label{sec:results}

Similarly to what happened with showing the causality of $\HB$ over
views, when establishing the causality of $\HB$ over phasers we
require an invariant that relates the various local views of a phaser.
While in the context of views, we must ensure that the wait phase does
not overtake the signal phase, in the context of phasers we must
ensure that its views may happen in parallel with each other, \ie
there must be no scheduling constraints within a phaser.

\begin{definition}[Well-ordered phaser]
  \label{def:wo}
  Let a \emph{well-ordered phaser} be such that $\Ph \PAR \Ph$.
  Let~$\PhWO$ be the set of all well-ordered phasers.
\end{definition}

Reduction \emph{cannot} introduce unsolvable scheduling constraints
within a phaser.

\begin{lemma}[Reduction preserves phaser well-orderedness]
  \label{lem:ph-wf-reduces}
  If $\Ph \in \PhWO$ and $\Ph \reduces \Phy$, then $\Phy \in \PhWO$.
\end{lemma}

We are now  ready to show that $\HB$ is a causality relation over phasers.

\begin{theorem}
  \label{lem:wo-pres-red}
  $(\HB,\PhWO)$ is a causality relation.
\end{theorem}

The execution of an HJ program must respect the scheduling restriction
imposed by Phase Ordering.
In the point of view of our formalization, the reduction relation
captures the execution of a single phaser operation.
Thus, \cref{thm:reduces-mhp} shows that the state after execution
cannot happen before the state before execution, or, in other words,
that the pre- and post-states of a phaser operation respect Phase
Ordering.

\begin{theorem}
  \label{thm:reduces-mhp}
  If $\Ph \in \PhWO$ and $\Ph \reduces \Phy$ we have that
  $\neg (\Phy \HB \Ph)$.
\end{theorem}

It is curious to consider that from $\Ph \reduces \Phy$ we can also
conclude $\neg (\Ph \HB \Phy)$.
Thus, it follows that.
\begin{lemma}
  \label{lem:red-par}
  
  If $\Ph \in \PhWO$, $\Ph \reduces \Phy$, then
  $\Ph \PAR \Phy$.
\end{lemma}

Be aware, however, that MHP does \emph{not} enjoy transitivity,
otherwise HB would be an empty relation!
\cref{ex:hb} is an evidence of when~$\Ph \PAR \Phy$ and
$\Phy \PAR \Phz$ but~$\neg (\Ph \PAR \Phz)$, as $\Ph \HB \Phz$.
This also tells us that for any two states~$\Ph$ and~$\Phy$ if we have
$\Ph \HB \Phy$, then state~$\Phy$ is the result of at least two phaser
operations.

The final theorem establishes that the execution of an HJ program
respects the scheduling restriction imposed by Phase Ordering.
While \cref{thm:reduces-mhp} relates the pre- and post-states of
executing a single operation, \cref{thm:hb-sred} generalizes this
result to any possible execution trace, showing that Phase Ordering
captures the execution order of instructions in programs that use
phasers.
Let~$\reduces^*$ be defined as the reflexive transitive closure of~$\reduces$.
%

\begin{theorem}[Absence of synchronization errors]
  \label{thm:hb-sred}
  $\Ph \in \PhWO $,
  $\Ph \reduces^* \Phy$,
  then
  $\neg (\Phy \HB \Ph)$.
\end{theorem}

Our results can be summarized into three groups.
The first group consists of \cref{lem:red-pres-wf,lem:ph-wf-reduces};
this serves as a steppingstone for our main results.
The former lemma establishes an invariant on a relationship between
wait and signal phases ($\ViewWF$), while the latter lemma establishes
an invariant on a relationship between any two views picked from a
state ($\PhWO$).
Since $\ViewWF$ and $\PhWO$ are preserved by our semantics, any
program that manipulates Habanero phasers can assume
\cref{lem:red-pres-wf,lem:ph-wf-reduces} to hold.
The second group consists of \cref{lem:hb-caus,lem:wo-pres-red} and
lets us relate views (and states) with the HB relation.
The third group consists of \cref{thm:reduces-mhp,thm:hb-sred} and it
lets us conclude that the execution of a program using Habanero
phasers respects the scheduling restriction imposed by Phase Ordering
(HB) relation.




%% file: conclusion.tex
\section{Conclusion}
\label{sec:conclusion}

In this paper we propose the first formalization of Habanero phaser
semantics and of Phase Ordering, from which we derive the
May-Happen-In-Parallel (MHP) and Happens-Before (HB) relations for
phaser operations, as part of an ongoing effort to formalize the
Habanero programming model.
Our definitions and proofs are mechanized using the Coq proof
assistant, it consists of 2600 lines of code and 140 lemmas.
Our next step is to verify data-race errors in parallel programs that
feature collective producer-consumer synchronization patterns.

\section*{Acknowledgments}

We thank Nick Vrvilo and the anonymous reviewers for their comments
and suggestions.

%% file: coq-results.tex
\section{Main definitions and results in Coq}
\newcommand{\CC}[1]{\ttfamily\color{comments}{#1}}

This section presents code listings, in Coq syntax, of the definitions
and results found in \cref{sec:phase-ordering,sec:results}.
The full proof scripts and auxiliary lemmas can be found online in our
open source project HJ-Coq~\cite{hj-coq}.
The next code listing shows definitions related to views.

\begin{coq}
 (* Module declares a name space to avoid avoid name collisions
    between the definitions of views and phasers. *)
 Module Taskview.
  (* (*@\CC{\cref{def:wf-views} for views, defined by three cases:}@*) *)
  Inductive Wellformed v : Prop :=
    | tv_wellformed_wait_cap_eq:
      WaitCap (mode v) ->
      wait_phase v = signal_phase v ->
      Wellformed v
    | tv_wellformed_wait_cap_succ:
      WaitCap (mode v) ->
      S (wait_phase v) = signal_phase v ->
      Wellformed v
    | tv_wellformed_so:
      mode v = SIGNAL_ONLY ->
      wait_phase v <= signal_phase v ->
      Wellformed v.
  (* (*@\CC{\cref{def:hb} for views:}@*) *)
  Inductive HappensBefore v1 v2 : Prop :=
    tv_hb_def:
      signal_phase v1 < wait_phase v2 ->
      SignalCap (mode v1) ->
      WaitCap (mode v2) ->
      HappensBefore v1 v2.
  (* (*@\CC{Declare the infix notation of the HappensBefore relation}@*) *)
  Infix "(*@$\HB$@*)" := HappensBefore : phaser_scope.
  (* (*@\CC{\cref{def:chb} for views, defined by three cases:}@*) *)
  Inductive CannotHappenBefore v1 v2 : Prop := 
    | tv_chb_ge:
       signal_phase v1 >= wait_phase v2 ->
       CannotHappenBefore v1 v2
    | tv_chb_so:
      mode v2 = SIGNAL_ONLY ->
      CannotHappenBefore v1 v2
    | tv_chb_wo:
      mode v1 = WAIT_ONLY ->
      CannotHappenBefore v1 v2.
  (* Define the infix notation of HappensBefore. *)
  Infix "(*@$\CHB$@*)" := CannotHappenBefore : phaser_scope.
End Taskview.
\end{coq}

Next, we list the definitions and notations related to phasers from
\cref{sec:phase-ordering}.

\begin{coq}
Module Phaser.
  (* (*@\CC{\cref{def:wf-views} for phasers:}@*) *)
  Inductive Wellformed (ph:phaser) : Prop :=
    ph_wellformed_def:
      (\forall t v, Map_TID.MapsTo t v ph -> Taskview.Wellformed v) ->
      Wellformed ph.
  (* (*@\CC{\cref{def:hb} for phasers:}@*) *)
  Inductive HappensBefore (ph1 ph2:phaser) : Prop :=
    ph_hb_def:
      \forall t1 t2 v1 v2,
      Map_TID.MapsTo t1 v1 ph1 ->
      Map_TID.MapsTo t2 v2 ph2 ->
      Taskview.HappensBefore v1 v2 ->
      HappensBefore ph1 ph2.
  (* Defines the infix notation of HappensBefore. *)
  Infix "(*@$\HB$@*)" := HappensBefore : phaser_scope.
  (* (*@\CC{\cref{def:chb} for phasers:}@*) *)
  Inductive CannotHappenBefore (ph1 ph2:phaser) : Prop := 
    ph_chb_def:
      (\forall t1 t2 v1 v2, Map_TID.MapsTo t1 v1 ph1 -> Map_TID.MapsTo t2 v2 ph2 ->
        Taskview.CannotHappenBefore v1 v2) ->
      CannotHappenBefore ph1 ph2.
  (* Defines the infix notation of CannotHappenBefore. *)
  Infix "(*@$\CHB$@*)" := CannotHappenBefore : phaser_scope.
  (* (*@\CC{\cref{def:wo}:}@*) *)
  Inductive WellOrdered x : Prop :=
    well_ordered_def: Facilitates x x -> WellOrdered x.
  (* (*@\CC{\cref{def:wo}:}@*) *)
  Inductive Par x y: Prop :=
    par_def: Facilitates x y -> Facilitates y x -> Par x y.
End Phaser.

\end{coq}
\newcommand{\CREF}[1]{\ttfamily\color{comments}\cref{#1}}%

Finally, we cross-reference the lemmas and theorems in the paper against
the Coq mechanization.
\begin{lstlisting}[language=Coq,literate=
    {\\forall}{{\color{dkgreen}{$\forall\;$}}}1
    {\\exists}{{$\exists\;$}}1
    {<-}{{$\leftarrow\;$}}1
    {=>}{{$\Rightarrow\;$}}1
    {==}{{\code{==}\;}}1
    {==>}{{\code{==>}\;}}1
%    {:>}{{\code{:>}\;}}1
    {->}{{$\rightarrow\;$}}1
    {<->}{{$\leftrightarrow\;$}}1
    {<==}{{$\leq\;$}}1
    {\#}{{$^\star$}}1 
    {\\o}{{$\circ\;$}}1 
    {\@}{{$\cdot$}}1 
    {\/\\}{{$\wedge\;$}}1
    {\\\/}{{$\vee\;$}}1
    {++}{{\code{++}}}1
    {~}{{$\neg$}}1
    {\@\@}{{$@$}}1
    {\\mapsto}{{$\mapsto\;$}}1
    {\\hline}{{\rule{\linewidth}{0.5pt}}}1
    {<}{$\HB$ }1 {>=}{$\CHB$}1 {->}{{$\rightarrow\;$}}1 {\\forall}{{\color{dkgreen}{$\forall\;$}}}1]
(* (*@\CREF{lem:red-pres-wf}@*) *)
Lemma ph_reduces_preserves_wellformed:
\forall ph t o ph', Wellformed ph -> Reduces ph t o ph' -> Wellformed ph'.
(* (*@\CREF{lem:hb-caus}@*) *)
Theorem tv_lt_trans: \forall x y z, Wellformed y -> x < y -> y < z -> x < z.
Theorem tv_lt_antisym:\forall x y,  Wellformed x -> Wellformed y -> x < y -> ~ (y < x).
Theorem tv_lt_irreflexive: \forall v, Wellformed v -> ~ (v < v).
(* (*@\CREF{lem:ph-wf-reduces}@*) *)
Lemma ph_hb_irreflexive: \forall ph, WellOrdered ph -> ~ (ph cc< ph).
Lemma ph_hb_antisym: \forall x y, WellOrdered x -> WellOrdered y -> x < y -> ~ (y < x).
Lemma ph_hb_trans: \forall ph1 ph2 ph3, WellOrdered ph2 -> ph1 < ph2 -> ph2 < ph3 -> ph1 < ph3
(* (*@\CREF{lem:wo-pres-red}@*) *)
Lemma reduces_ne: \forall x y, WellOrdered x -> SReduces x y -> x >= y.
(* (*@\CREF{lem:red-par}@*) *)
Lemma reduces_par: \forall x y, Wellformed x -> WellOrdered x -> SReduces x y -> x || y.
(* (*@\CREF{thm:reduces-mhp}@*) *)
Lemma ph_ge_reduce:
\forall ph t o ph', Wellformed ph -> WellOrdered ph -> Reduces ph t o ph' -> ph' >= ph.
(* (*@\CREF{thm:hb-sred}@*) *)
Lemma ph_s_reduces_trans_refl_ge: \forall x y, Wellformed x -> WellOrdered x -> 
clos_refl_trans phaser SReduces x y -> y >= x.
\end{lstlisting}

%% file: proofs.tex
\section{Proof sketches}
\label{sec:proofs}

The proofs for all lemmas and theorems in this paper are machine checked in~\cite{hj-coq}.
In this section, we show the proof sketches for the main results.

\paragraph{\cref{lem:wo-pres-red}}
$(\HB,\PhWO)$ is a causality relation.
\begin{proof}
  $(\HB,\PhWO)$ is transitive: if $\Ph \HB \Phy$ and~$\Phy \HB \Phz$,
  then $\Ph \HB \Phz$.
  We invert~$\Ph \HB \Phy$ and get that there exists~$\Tid_1$
  and~$\Tid_2$ such that $\Ph(\Tid_1) = \View_1$,
  $\Phy(\Tid_2) = \View_2$, and $\View_1 \HB \View_2$.
  Similarly, we invert~$\Phy \HB \Phz$ and get that there exists
  $\Tid_3$ and~$\Tid_4$ such that $\Phy(\Tid_3) = \View_3$,
  $\Phz(\Tid_4) = \View_4$ such that $\View_3 \HB \View_4$.
  From~$\Phy \in \PhWO$, $\Phy(\Tid_3) = \View_3$, and
  $\Phy(\Tid_2) = \View_2$, thus~$\View_3 \CHB \View_2$.
  From $\View_1 \HB \View_2$, $\View_3 \HB \View_4$, and
  $\View_3 \CHB \View_2$, we can conclude that $\View_1 \HB \View_4$,
  and therefore $\Ph \HB \Phz$.
  
  $(\HB,\PhWO)$ is irreflexive: if $\Ph \in \PhWO$ then $\neg (\Ph \HB \Ph)$.
  From $\Ph \in \PhWO$ we get that $\Ph \CHB \Ph$.
  Then, we apply \cref{rem:conv} and get that $\neg (\Ph \HB \Ph)$.
  
  $(\HB,\PhWO)$ is asymmetric: if $\Ph \HB \Phy$, then
  $\neg (\Phy \HB \Ph)$.
  Using \cref{rem:conv} it is enough to show that $\Phy \CHB \Ph$,
  specifically that if $\Phy(\Tid_1) = \View_1$ and
  $\Ph(\Tid_2) = \View_2$, then $\View_1 \CHB \View_2$.
  It can be shown that for any pair of views we have that
  $\View_1 \HB \View_2$ or $\View_1 \CHB \View_2$.
  Since the latter concludes the proof directly, we proceed to show
  that the former case,~$\View_1 \HB \View_2$, leads to a contradiction.
  From $\Ph \HB \Phy$ we have that there exists task~$\Tid$
  and~$\Tid'$ such that $\Ph(\Tid) = \View$, $\Phy(\Tid')=\Viewy$, and
  $\View\HB \Viewy$.
  The contradiction arises from arriving at
  $\neg (\View_1 \CHB \Viewy)$ and $\View_1 \CHB \Viewy$.
  First, we get $\View_1 \CHB \Viewy$ by applying \cref{def:wo} to
  $\Ph \in \PhWO$, $\Phy(\Tid_1) = \View_1$, and $\Phy(\Tid')=\Viewy$.
  Second, we show $\neg (\View_1 \CHB \Viewy)$.
  Applying \cref{def:wo} to $\Ph \in \PhWO$, $\Ph(\Tid) = \View$, and
  $\Ph(\Tid_2) = \View_2$ results in~$\View \CHB \View_2$.
  As we have seen in the proof of transitivity, from
  $\View_1 \HB \View_2$, $\View \HB \Viewy$, and $\View \CHB \View_2$,
  we conclude $\View_1 \HB \Viewy$.
  Applying \cref{rem:conv} we get that~$\neg (\View_1 \CHB \Viewy)$,
  which leads to the contradiction.
\end{proof}

\paragraph{\cref{thm:reduces-mhp}}
  If $\Ph \in \PhWO$ and $\Ph \reduces \Phy$ we have that
  $\neg (\Phy \HB \Ph)$.

\begin{proof}
  By inverting the hypothesis $\Ph \reduces \Phy$ we get
  $\Ph \Reduces\Op \Phy$.
  Next, we invert $\Ph \Reduces\Op \Phy$ and obtain four cases, one
  for each constructor of~$\Op$.
  For each case it is enough to show that given~$\Ph(x) = \View_x$ and
  $\Phy(y) = \View_y$, we can obtain $\View_y \CHB \View_x$.
  
  Cases~$\Op =\Osignal$ and $\Op = \Owait$ proceed similarly.
  We test if~$y = \Tid$.
  If $y \neq \Tid$,
  then $\Phy(y) = \Ph(y)$.
  We can obtain $\View_y \CHB \View_x$ from $\Phy(y) = \View_y$ ,
  $\Phy(y)=\View_y$ and $\Phy \in \PhWO$ (which we get from
  \cref{lem:ph-wf-reduces}, $\Ph \reduces \Phy$, and $\Ph \in \PhWO$).
  Otherwise $y = \Tid$, and we get that there exists a view~$\View$
  such that $\Ph(y)=\View$.
  Let the increment of the signal phase (wait phase) be denoted
  by~$\Op(\View)$ where~$\Op(\View) = \View_y$.
  At this point, we have $\Ph(y) = \View$ 
  and~$\Phy(y) = \Op(\View)$.
  Then, we just need to show that $\Op(\View_x) \CHB \View_x$, given
  that $\View \CHB \View_x$, which we get from $\Ph(y)=\View$,
  $\Ph(x)=\View_x$, and $\Ph\in \PhWO$.
  
  Case~$\Op = \Oregister[\Tidy] \Regmode$, where we
  have~$\Ph(\Tid) = \View$.
  We test if $\Tidy = y$.
  If $\Tidy \neq y$, then $\Ph(y) = \View_y$.
  Thus, we have $\View_y \CHB \View_x$ from $\Ph(y) = \View_y$,
  $\Ph(x) = \View_x$, and $\Ph \in \PhWO$.
  Otherwise, $\Tidy = y$, and
  therefore~$\View_y = (\Mode \assign \Regmode)$.
  From $\Ph(\Tid) = \View$, $\Ph(x) = \View_x$, and $\Ph \in \PhWO$,
  we conclude $\View \CHB \View_x$.
  We close the case by showing that from~$\View \CHB \View_x$ then
  $(\Mode \assign \Regmode) \CHB \View_x$ holds.
  
  Case~$\Op = \Odrop$, since we have $\Phy = \Ph - \Tid$, then
  $\Ph(y) =\View_y$.
  From $\Ph(y) = \View_y$, $\Ph(x) = \View_x$, and $\Ph \in \PhWO$,
  we get $\View_y \CHB \View_x$.
\end{proof}

\paragraph{\cref{thm:hb-sred}}
  $\Ph \in \PhWO $,
  $\Ph \reduces^* \Phy$,
  then
  $\neg (\Phy \HB \Ph)$.

\begin{proof}
  We state our result in terms of $\Phy \CHB \Ph$ and, but we can use
  \cref{rem:conv} to obtain $\neg (\Phy \HB \Ph)$.
  The proof follows by induction on the derivation tree of
  $\Ph \reduces^* \Phy$.
  For the base case we have that~$\Phy = \Ph$, and we show
  that~$\Ph \CHB \Ph$ holds.
  
  For the inductive case we have that there exists a phaser~$\Phz$
  such that $\Ph \reduces^* \Phz$, $\Phz \reduces \Phy$, and
  $\Phz \CHB \Ph$.
  From $\Ph \in \PhWF$ (which states that every view~$\View$ in~$\Phz$
  is~$\View \in \ViewWF$) and $\Ph \reduces^* \Phz$ we can get that
  $\Phz \in \PhWF$, by performing induction on the structure of
  $\Ph \reduces^* \Phz$ and using Lemma~\ref{lem:red-pres-wf}.
  Similarly, we get that $\Phz \in \PhWO$, by performing induction on
  the structure of $\Ph \reduces^* \Phz$ and using
  Lemma~\ref{thm:reduces-mhp}.
  The final step of the proof is to show that if $\Phz \in \PhWF$
  $ \Phz \CHB \Ph$, and $\Phz \reduces \Phy$, then $\Phy \CHB \Ph$.
  At this point, we perform a case analysis in the reduction
  relation~$\Phz \reduces \Phy$.
  The case for $\Phz \Rdrop \Phy$ is trivial, thus we shift our
  attention to the remaining cases, where
  $\dom \Phz \subseteq \dom \Phy$, and the crux of the proof is
  showing that if~$\Ph(x) = \View_x$ and~$\Phz(y) = \View_y$, and
  $\Phy(z) = \View_z$, then
  $\View_z \CHB \View_x$.
  
  We now check if~$t = z$.
  If the check succeeds, then we can conclude that there is a
  view~$\View$ such that $\Phz(z)= \View$ such that $\View_x$ results
  by applying a signal or wait to~$\View$, notation
  $\View_x=\Op(\View)$.
  Now, because we have that~$\Phz \CHB \Ph$, then
  $\View \CHB \View_x$, so we just need to show that
  $\Op(\View) \CHB \View_x$, which we omit detailing.
  
  Finally, we address the case where~$t \neq z$.
  We inspect~$\Op$ and discuss the non-trivial case, when there exist
  $\Tid'$, $\Regmode$, and $\View$ such
  that~$\Op = \Oregister[\Tid'] \Regmode$, 
  $\Phz(\Tid) = \View$, and $\View_z = (\Mode \assign \Regmode)$.
  Recall, we want to show that $(\Mode \assign \Regmode) \CHB \View_x$
  holds.
  The proof can be concluded using three premises:
  $ \CanWait [\Regmode] \implies \CanWait$ and
  $\CanSignal [\Regmode] \implies \CanSignal$, which we get from the
  reduction rule on~$\Oregister [\Tid'] \Regmode$; and
  $\View \CHB \View_x$, which we get from $\Phz \CHB \Ph$.
\end{proof}